\documentclass[two column, superscriptaddress, aps]{revtex4-2}
\usepackage{graphicx, amsmath,xcolor}

\begin{document}
	
	\title{Zonal flows and reversals of cortically confined active suspensions}
	
	\author{J.~S.~Yodh}
	\affiliation{Department of Physics, Harvard University, Cambridge, Massachusetts 02138}
	\author{F.~Giardina}
	\affiliation{John A. Paulson School of Engineering and Applied Sciences, Harvard University, Cambridge, Massachusetts 02138}
	\author{S.~Gokhale}
	\affiliation{Physics of Living Systems, Department of Physics, Massachusetts Institute of Technology, Cambridge, MA 02139}
	\author{L.~Mahadevan}
	\affiliation{Department of Physics, Harvard University, Cambridge, Massachusetts 02138}
	\affiliation{John A. Paulson School of Engineering and Applied Sciences, Harvard University, Cambridge, Massachusetts 02138}
	\affiliation{Department of Organismic and Evolutionary Biology, Harvard University, Cambridge, Massachusetts 02138}

    \begin{abstract}
At sufficiently high concentrations, motile bacteria suspended in fluids exhibit a range of ordered and disordered collective motions.  Here we explore the combined effects of confinement, periodicity and curvature induced by the active motion of \textit{E. coli} bacteria in a thin spherical shell (cortex) of an oil-water-oil (O/B/O) double emulsion drop.  Confocal microscopy of the bacterial flow fields shows that at high density and activity, they exhibit azimuthal zonal flows which oscillate between counterclockwise and clockwise circulating states. We characterize these oscillatory patterns  via their Fourier spectra and the distributions of their circulation persistence times. To explain our observations, we used numerical simulations of active particles and characterize the two-dimensional phase space of bacterial packing fraction and activity associated with persistent collective motions. All together, our study shows how geometric effects lead to new types of collective dynamics. 
    \end{abstract}
 
    \maketitle
	
Collections of individual agents that convert internal energy into mechanical work can yield complex patterns in space-time and have been studied in a wide variety of contexts using both synthetic particles and living organisms \cite{bechinger2016active, marchetti2013hydrodynamics, ramaswamy2010mechanics}. A particularly interesting class of active systems amenable to a range of experimental manipulations are motile bacteria that exhibit nontrivial collective flow patterns \cite{alert2021active,wensink2012meso, dunkel2013fluid,wioland2013confinement, lushi2014fluid, liu2021viscoelastic, hamby2018swimming} that range from simple oscillations to turbulence. While these and related experiments with synthetic active matter \cite{zhang2021polar, zhang2022guiding, bricard2013emergence, geyer2019freezing, palacci2013living, sanchez2012spontaneous, narayan2007long} 
have provided much inspiration for the development of a vast number of theoretical models, relatively little is known about collective flows in 3D confined geometries.  

Here we explore the behavior of a dense suspension of motile bacteria moving on thin spherical shells to explore the role of confinement, curvature and domain periodicity. While this geometry has been studied in theory and simulation \cite{shankar2017topological, mickelin2018anomalous, sknepnek2015active}, experimental work has been limited \cite{keber2014topology, hsu2022activity}. 
Our polar active system consists of highly motile bacteria with an aspect ratio ($a \sim 3$) that move approximately ten body lengths per second. Critically, because bacteria can move in the polar and the azimuthal directions, we observe using confocal microscopy persistent zonal flows which switch between counterclockwise and clockwise circulating states.  These transitions occur when populations of bacteria moving in the counterclockwise direction are replaced with those moving in the clockwise direction.  Presumably, these events are caused by relatively slow precession of bacterial flow structures within the shell.  
    \begin{figure*}
	\centering
	\includegraphics[scale = 1]{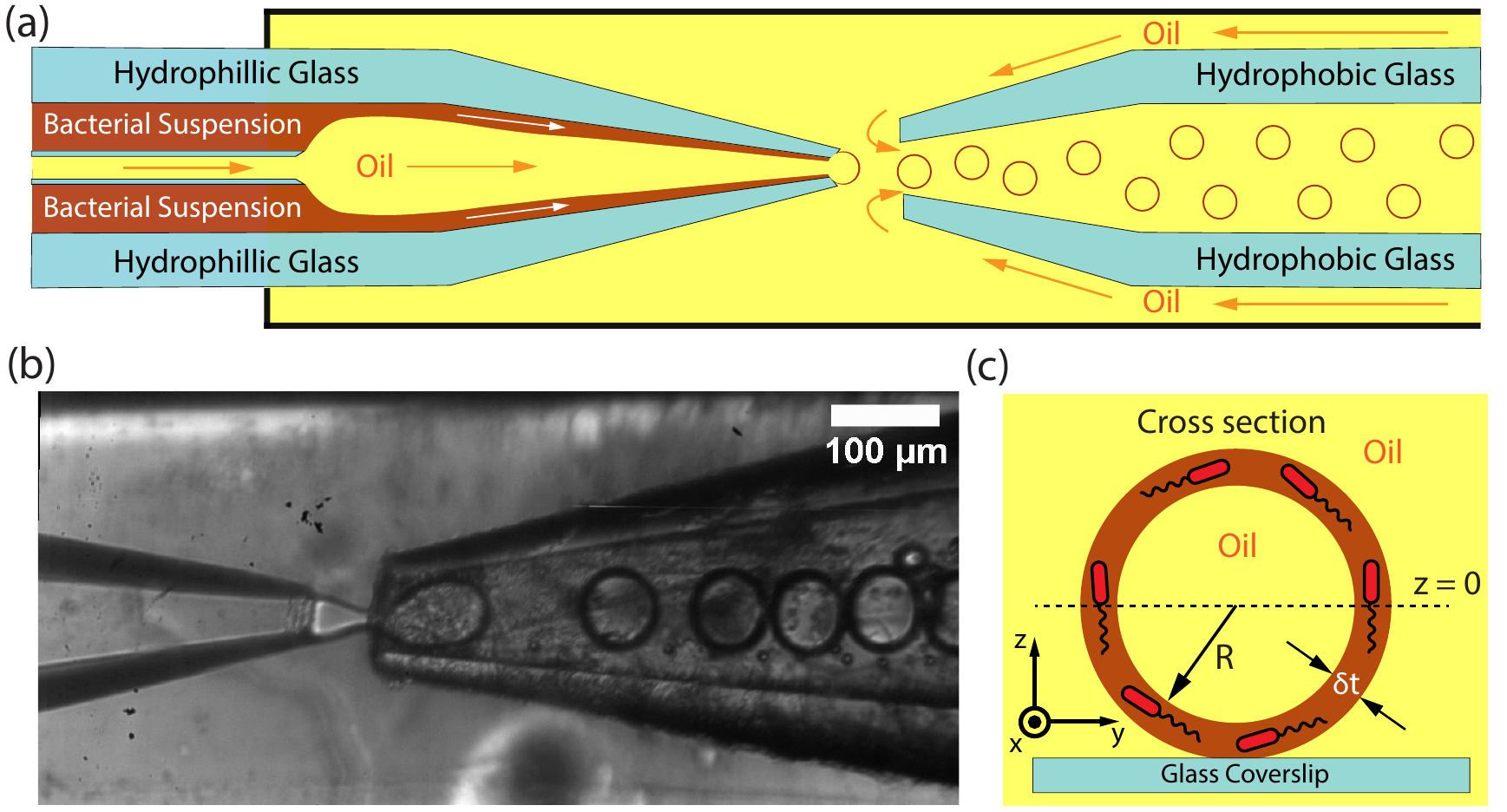}
	   \caption{(a) Schematic of the microfluidic device used to generate thin-shelled oil-bacterial suspension-oil double emulsions.  The flow rates of the two upstream lines are order 100 $\mu$L/h, and the flow rate of the downstream oil line is approximately 300 $\mu$L/h.  (b) Micrograph of break-off of a bacterial double emulsion approximately 60 $\mu$m in diameter.  Note, if the device had generated single emulsions, then the droplets would appear entirely dark. (c) Schematic of the cross section of a double emulsion drop (not to scale) filled with bacteria of average length $\ell = 3$ $\mu$m.  The double emulsion drop is characterized by its radius, $R$, which varies between 5 and 49 $\mu$m, and its shell thickness, $\delta t$, which ranges between 1 and 3 $\mu$m.  The shell thickness is roughly the width of one bacteria so that the system quasi-2D. The volume fraction of bacteria within the shell is near random close packing in 2D, $\varphi \sim 0.7-0.8$}
    \end{figure*}	
The bacteria preparation protocol is similar to that described in \cite{schwarz2016escherichia} (see Supplemental Material - SM - for further details of our preparation procedure).  Briefly, the experiment uses an RP437 derivative of \textit{E. coli} with $\Delta$CheY mutation and red fluorescent plasmid (pSBIK3-RFP) with Kanamycin resistance. The $\Delta$CheY mutation inhibits bacteria from tumbling \cite{scharf1998control}, the pSBIK3-RFP enables fluorescence, and the Kanamycin resistance is needed to kill off bacteria without the fluorescent plasmid (\textit{e.g.}, which could form by mutation). In isolation, these smooth swimmers still change direction via rotational diffusion; at higher concentrations, the smooth swimmers also change direction via collisions.  Single colonies of \textit{E. coli} are picked from agar plates and are cultured overnight in Luria Broth (LB) to saturation.  The \textit{E. coli} are then diluted 1:100 and grown in Tryptone Broth (TB) for $\sim$5 hours.  They are then concentrated $\sim$100x in a TB solution with 1\% by weight F108 Pluronic, a surfactant which aids formation and stability of the double emulsions. 

 To confine the bacteria in a thin spherical shell, we created thin-shelled oil-bacterial-oil (O/B/O) double emulsions using microfluidics. We utilized a coaxial glass capillary flow focusing device \cite{kim2011double} schematically shown in Fig.~1(a) and optically shown Fig.~1(b), comprised of two coaxially aligned, circular, tapered glass capillaries inside a square glass capillary.  Glass in the injection capillary is hydrophilic, and glass in the collection capillary is hydrophobic.  A third narrower injection capillary is threaded into the injection capillary, and facilitates the formation of thin-shelled double emulsions. The inner phase of the double-emulsion, N-Hexadecane with 5\% Span 80  by weight chosen for its viscosity being close to that of the aqueous phase,  was injected into the narrowest and farthest upstream injection capillary, while the capillary downstream of this inner capillary carried the \textit{E. coli}-surfactant (Pluoronic F108, 1\% by weight) suspension along the hydrophillic glass wall.  Finally, N-Hexadecane with 5\% by weight Span 80 was injected into the capillary tube farthest downstream.  The Oil/Bacterial/Oil mixture breaks into highly mono-disperse droplets forming a double-emulsion via the Rayleigh-Plateau instability  [see Fig.~1(b) and Supplementary Video 1] with radii roughly set by the injection capillary diameter (which we vary between $5-50 \mu m$), though  other factors such as flow rate also influence drop size.  Critically, we found that slower flow rates promote formation of robust double emulsions (See SM for further details).

The double emulsion drops are then plated onto a glass coverslip and imaged using spinning disk confocal microscopy. This yields a series of planar-slice videos at distinct cross-sections, each corresponding to an annular disk of thickness corresponding approximately to the length $\ell\sim 3$ $\mu$m of a single bacterium. We define the equatorial plane as $z = 0$ with the z-axis anti-parallel to the gravity and the $x$- and $y$-axes lying in the plane transverse to it,  schematically shown in Fig.~1(c), with a corresponding optical micrograph in Fig.~2(a).  To measure the mean bacterial flows in the droplet (see Supplemental Video 2,3 and Fig.~S3), we track the angular displacement of intensity as a function of time, in the vicinity of the equator $z=0$) using an angular particle-image-velocimetry (PIV) algorithm which computes intensity-intensity correlations in time, taking care to eliminate drift. 

    \begin{figure*}
	\centering
	\includegraphics[scale = 1]{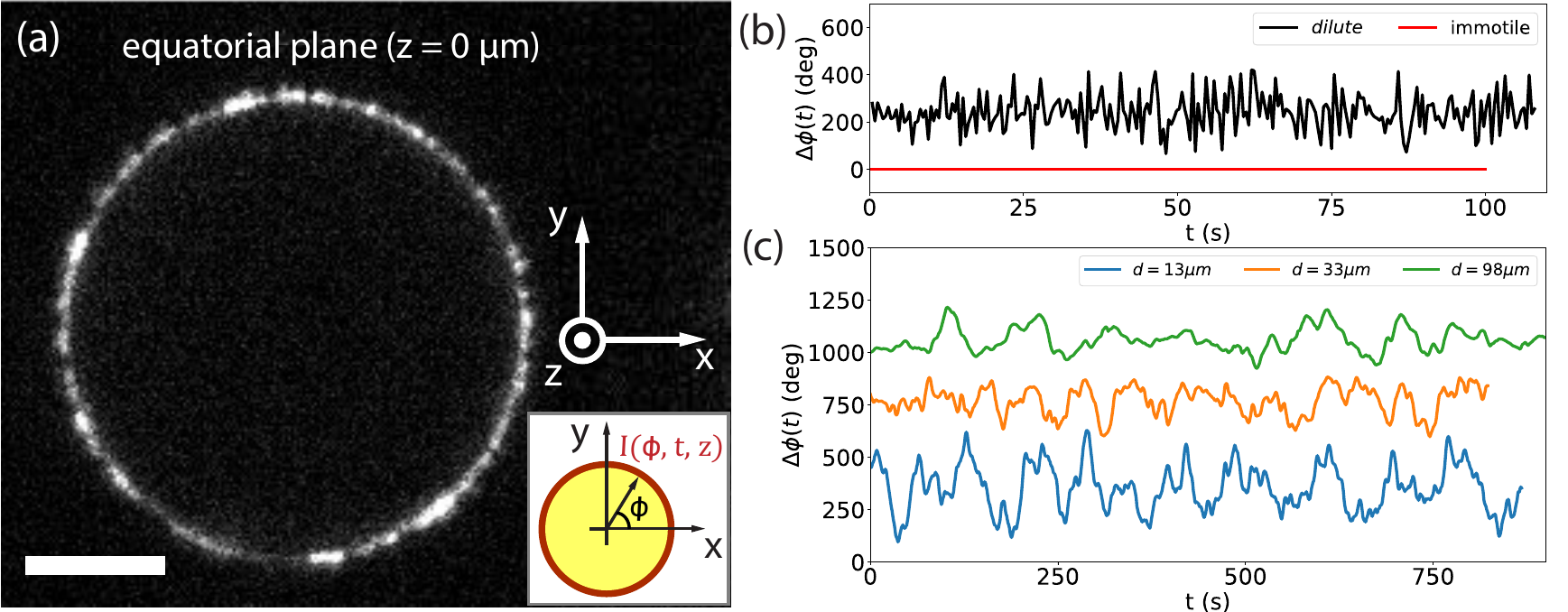}
    	\caption{(a) Equatorial planar cross section ($z = 0$) from a confocal stack of a 33 $\mu$m diameter double emulsion drop.  The $z-$axis points out of the page (in contrast to the schematic in Fig.~1(c)). Scale bar is 10 $\mu$m.  Inset: schematic of the drop cross section and its coordinate system; $I(\phi, t, z)$ represents the continuous spatiotemporal fluorescence intensity within the shell around the circumference.   (b) Representative angular displacements as a function time, $\Delta \phi (t)$, for the equatorial plane of two different double emulsion drop. One double emulsion drop contained a dense suspension of immotile bacteria (red). The other double emulsion drop contained a relatively dilute suspension of active bacteria (black) at packing fraction $\varphi \sim 0.25$. The vertical axis is offset for visualization. (c) Same as (b) for three size double emulsions with active bacteria: $d = 13$ $\mu$m (small), $d = 33$ $\mu$m (medium), and $d = 98$ $\mu$m (large). The vertical axis is offset for visualization.}
    \end{figure*}

In Fig.~2(b), we show the angular trajectories of intensity $\Delta\phi(t)$, as a function of time for a double emulsion drops filled with immotile bacteria (as a control) and filled with a relative dilute suspension of active bacteria at packing fraction $\varphi \sim 0.25$. Immotile bacteria do not exhibit constant circulation in one direction or oscillatory motion, while the motile bacteria show fluctuating angular trajectories in time, sometimes circulating in a clockwise direction, and then a counterclockwise direction; they do not circulate in a single direction in contrast to recent theory/simulation predictions \cite{shankar2017topological,sknepnek2015active,bruss2017curvature}, or experiments with low-activity synthetic polar suspensions \cite{hsu2022activity}.   In Fig.~2(c), trajectories of intensity for dense motile suspensions in three representative double emulsion drops with diameters $d \in[13, 33, 98]$ $\mu$m, offset vertically for clarity show sustained bouts of counterclockwise and clockwise motion combined with switching (see Supplemental Video 2).  We first analyzed the angular trajectories using their autocorrelations.  For all double emulsion sizes, the autocorrelation function de-correlated completely in $t\in[15,30]$ seconds; however, there was no clear size dependent signature in the data (see SM for details). \
 
Fig.~3(a) shows representative (discrete) fast Fourier transforms (FFTs) of the measured angular displacement trajectories for small, medium, and large double emulsion drops computed explicitly using $\hat{F}(k) = \sum_{j=1}^{n}\Delta \phi(j) W_n^{(j-1)(k-1)}$ where $W_n = e^{-2\pi i/n}$, $j$ indexes the vector $\Delta \phi$, and $k$ indexes the frequency, $f$, of the FFT. The FFTs for all droplets exhibit roughly the same spectral structure, with most activity in the range of $\sim 0.01-0.03$ Hz; outside of this low-frequency band, the FFT amplitude decays rapidly with increasing frequency.  The fluctuations connected with this low-frequency experience periodic changes in circulation direction (counterclockwise/clockwise) on timescales between $\sim 33 - 100$ seconds. The mean/peak frequency of the band and its full-width-half-max (FWHM) are summarized in the inset of Fig.~3(a).   
	
	\begin{figure}
		\centering
 		\includegraphics[scale = 1]{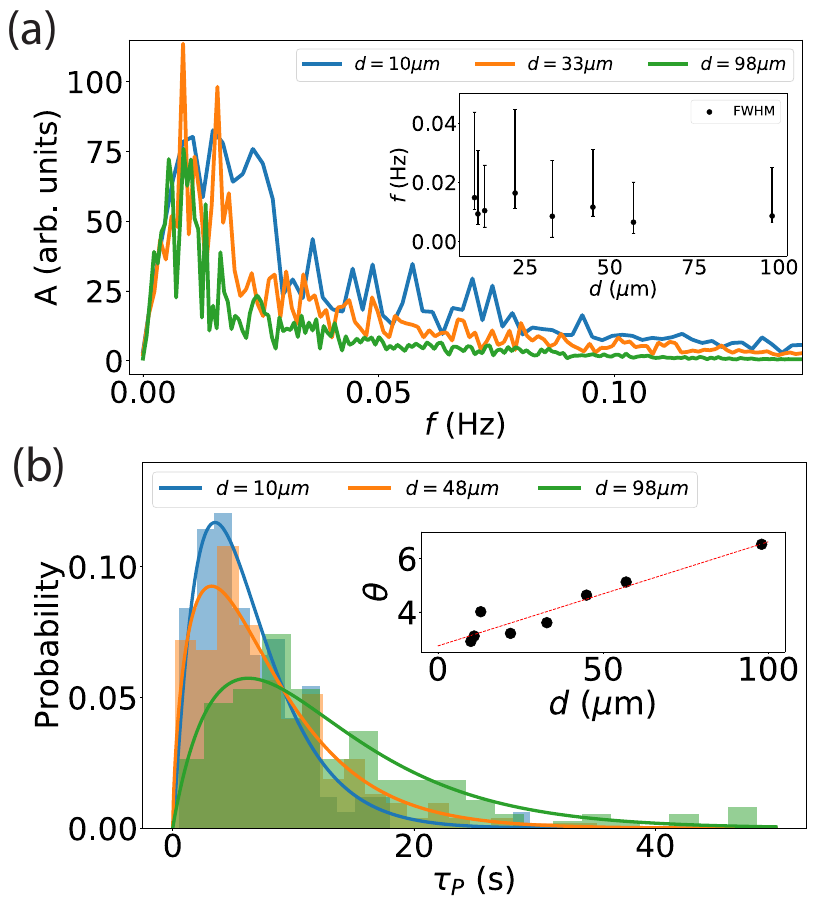}
		\caption{(a) Representative Fourier spectra of angular displacement trajectories for small, medium, and large double emulsion drops.  Inset: Mean/peak Full-width-half-max (FWHM) frequencies as a function of double emulsion diameter.  (b)  Histograms of persistent (counter)clockwise azimuthal flows and accompanying $\Gamma$ distribution fits, defined as $f(\tau_P, k, \theta) = \frac{\tau_P^{k-1}e^{-\tau_P/\theta}}{\theta^k \Gamma(k)}$, for small, medium, and large double emulsion drops. Insets (top to bottom): scale parameter, $\theta$, from $\Gamma$ distribution fits of  as a function of droplet diameter with a shown best-fit $\theta = 0.03d + 2.7$.}
	\end{figure}	


To further characterize the azimuthal flows, we measure the probability distribution of persistence times, $\tau_P$, defined as the time period (or residence time) between clockwise/counterclockwise circulation bouts, with switching characterized by a zero-crossing in the derivative of the angular trajectories found in Fig.~2(b) (see SM for example). Note, $\phi(t)$ can rapidly switch from counter-clockwise to clockwise and back over a timescale comparable to the volumetric confocal scan speed.  We attempt to filter our such rapid switching using a two point box smooth.  In Fig.~3(b) we exhibit these probability distribution histograms that are well fit by a $\Gamma$ distribution of the form $f(\tau_P, k, \theta) = \frac{\tau_P^{k-1}e^{-\tau_P/\theta}}{\theta^k \Gamma(k)}$, with $\tau_P ,k,\theta >0$. We observe that the probability distributions have a conserved shape parameter centered around $k = 1.89$; however, the  scale parameter varies linearly with the diameter, $d$, of the double emulsion according to $\theta (d) = 0.03d + 2.7$.  In the inset of Fig.~3(b), we show the scale parameter as a function of the droplet diameter; our data suggest that larger double emulsions can support persistent (counter)clockwise motion for longer $\tau_P$. 
     
     \begin{figure*}
		\centering
		\includegraphics[scale = 1]{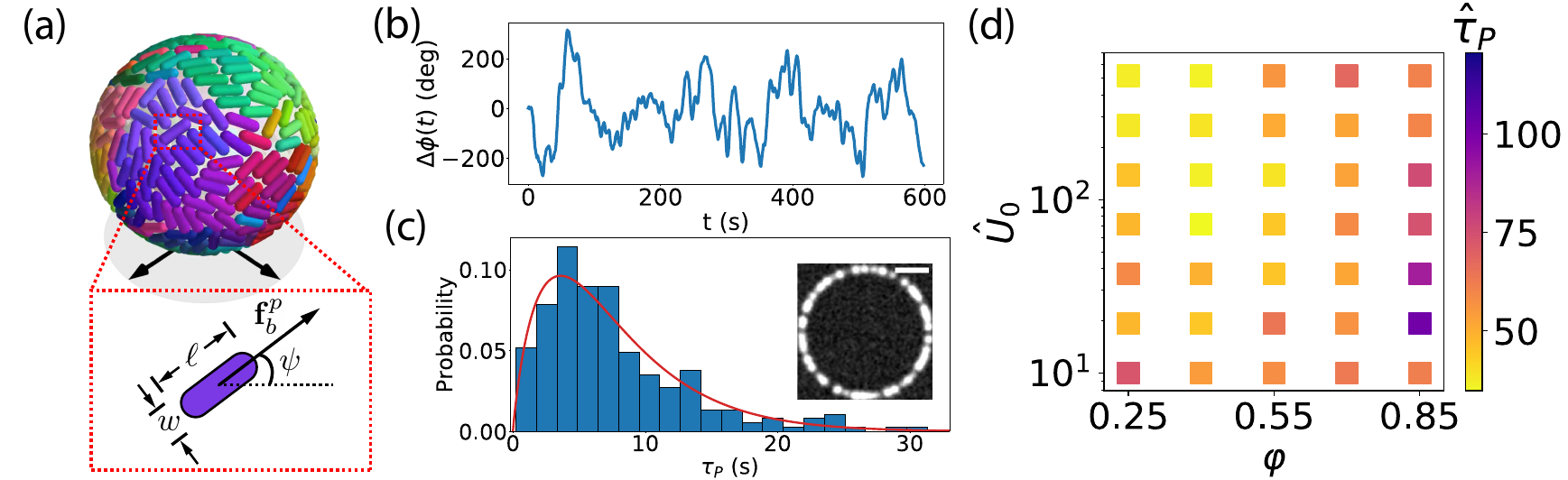}
		\caption{(a) Visualization of simulated self-propelled rods at packing fraction $\varphi = 0.84$ constrained to an oblate spheroid of semi-major axis $a = 10$ $\mu$m $\simeq b$ .  The dynamics of an individual rod are governed by equations (\ref{force}) and (\ref{torque}). Color indicates the direction of the velocity of an individual rod.  A self-propelled rod is schematically represented below.  (b) Representative angular trajectory as a function time, $\Delta \phi (t)$, for the equatorial plane derived from simulated rods.  The angular trajectories were obtained using the same method used for experimental data. (c) Normalized probability distribution of persistence times derived from equatorial cross sections of simulated data from six separate ten minute simulations with accompanying $\Gamma$ distribution fit. Inset: simulated equatorial cross section.  Scale bar is 5 $\mu$m. (d) Phase space for average persistence times in the $\varphi$-$\hat{U}_0$ plane, where $\hat{U}_0$ is the nondimensionalized ratio of Yukawa amplitude to scaled propulsive force and $\varphi$ is the packing fraction.  All units are dimensionless.  Note that $\hat{\tau}_P$ is  calculated using the total angular momentum of the entire system whereas in Fig.~3(b) and Fig.~4(c) $\tau_P$ is computed using only the equatorial plane.}
    \end{figure*}
    

To understand our experiments, we employ numerical simulations of active agents confined to a spheroidal domain of given shape, varying their packing fraction and activity to delineate their collective behavior (see SM for details).  Each of the $N$ identical self-propelled rods of length $\ell$, width $w$, and aspect ratio $\alpha = \ell/w = 3$ is confined to an oblate spheroid with semi-major axes, $a$, $a$, and $b$, such that $b/a = 0.95$ (consistent with experiment), and modeled by three distinct spherical segments in series.  Individual rod overdamped dynamics are determined by local forces and torques generated by self-propulsion, interactive forces, and drag  described by the equations for their position $\mathbf{r}(t)$ on the spheroid and polar orientation $\psi$ measured in the local tangent plane of the particle, 
    
    \begin{eqnarray}
        D_T \frac{d\mathbf{r}}{dt} &=& \mathbf{f}^p_b -\nabla_r U(r) \label{force},\\
        D_R\frac{d\psi}{dt} &=& - \nabla_\psi U(r). \label{torque}
    \end{eqnarray}
where $\mathbf{f}^p_b = D_T v_b \mathbf{e}_p$ is a constant propulsive force  along the vector, $\mathbf{e}_p$,  $v_b = 20$ $\mu$m/s is the speed of a bacteria freely swimming, and $D_T$ and $D_R$ are the Stokes drag coefficients for an ellipsoid, and $U(x) \sim e^{-x/\lambda}/x$ is a repulsive Yukawa potential between rods separated by a distance $x$ and $\lambda = w$ is a screening length \cite{wensink2012meso, wensink2012emergent}.  This short-ranged repulsive potential ensures that collisions between neighboring rods induces nematic alignment between rods and that rods remain appropriately separated.  In addition, it prevents the formation of giant islands of jammed rods and accompanying voids, which we do not observe in experiment. We note that the equations \ref{force} and \ref{torque} are projected using the local Jacobian to account for the dynamical constraints enforced by the spheroidal surface (see SM for details).

\begin{figure}[h!]
		\centering
		\includegraphics[scale = 1.0]{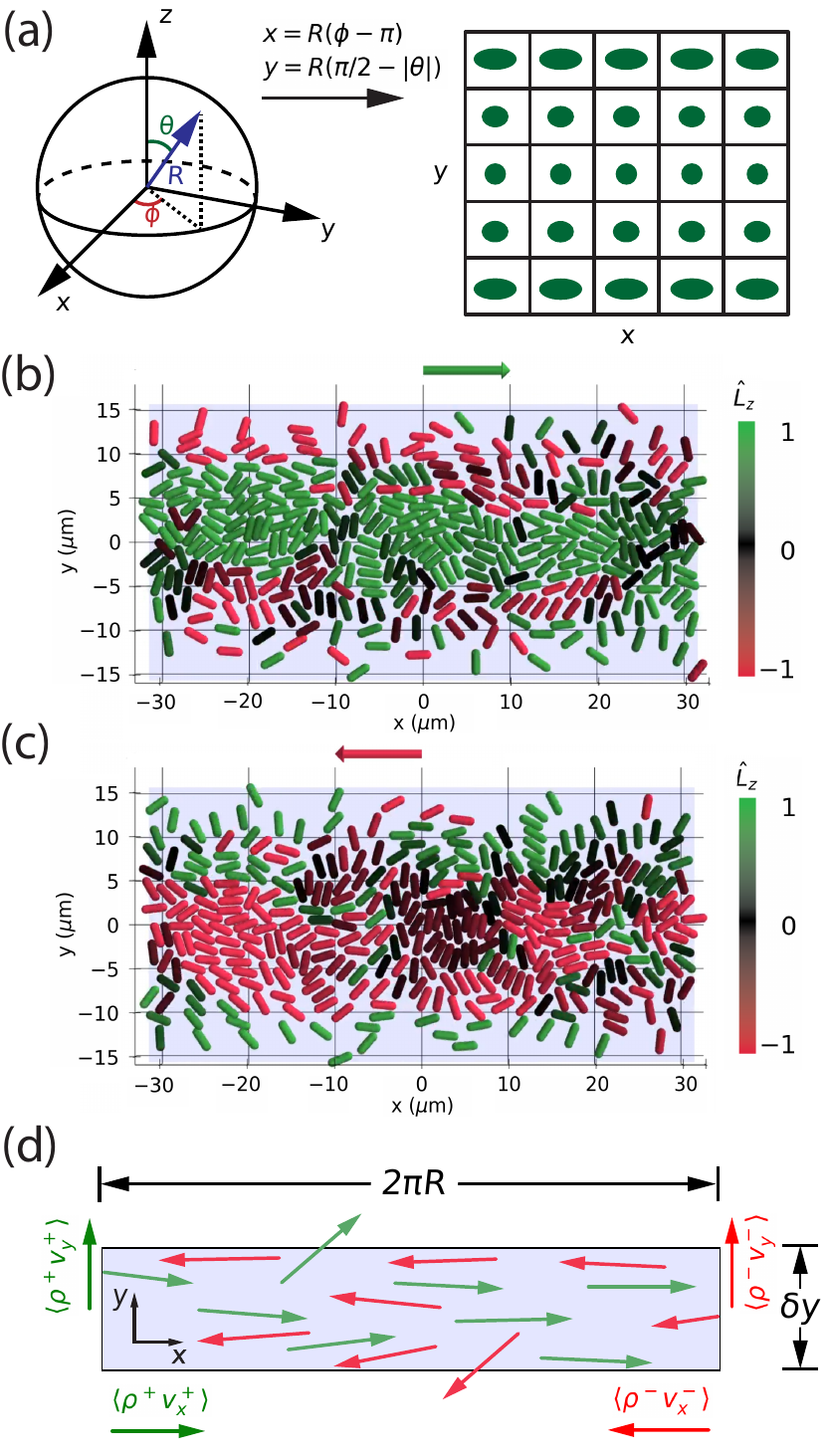}
		\caption{(a) Equirectangular stereographic projection schematic.  The map takes azimuthal and polar coordinates $\phi$ and $\theta$ respectively to $x$ and $y$ coordinates according to $x = R(\phi - \pi)$ and $y = R(\pi/2 -|\theta|)$ where we approximate the semi major and minor axes of the spheroid to be $R$. (b)  Simulated self-propelled rods on a 20 $\mu$m diameter spheroid with persistent counterclockwise flow. Color scale denotes the normalized angular momentum of an individual rod projected onto the $z$ axis. (c) Same as (b) with persistent clockwise flow.   (d) Schematic of stereographic band of length $2\pi R$ and width $\delta y$.  Green arrows denote populations of rods moving with positive angular momentum and red arrows denote populations of rods moving with negative angular momentum. The rods can be coarse grained into average fluxes moving in the $x$ and $y$ directions with $\langle \rho^+v^+_x\rangle$ and $\langle \rho^-v^-_x\rangle$ representing green and red flux in the $\theta$ direction respectively and $\langle \rho^+v^+_y\rangle$ and $\langle \rho^-v^-_y\rangle$ representing green and red flux in the $y$ direction respectively.}
    \end{figure}
    
With $\lambda = w$, there are five relevant non-dimensional parameters: the ratio of the major and minor axes of the spheroid to the Yukawa screening length, $\hat{a} = a/\lambda$, $\hat{b} = b/\lambda$, the rod aspect ratio, $\alpha = \ell/\lambda$, the packing fraction $\varphi$, and the ratio of Yukawa amplitude to the scaled propulsive force,  $\hat{U}_0 = U_0/|\mathbf{f}^p_b|\lambda^2$. Shown in Fig.~4(a) is a representative snapshot of the simulation (see SM Video 3 for simulations at different normalized activity $\hat{U}_0$). Here, self-propelled rods of length $\ell = 3$ $\mu$m are confined to an oblate spheroid of major and minor axis $a \sim b \sim 10$ $\mu$m, values chosen to match experiments. The color in Fig.~4(a) corresponds to rod velocity; note the velocity of an individual rod is not necessarily collinear with its orientation.

To compare our results with experiments, we apply the same angular PIV algorithm and data analysis to the equatorial cross-sections from simulated data after processing them  through the same filters as the experimental data (see Supplemental Video 3).  Shown in Fig.~4(b) is a representative angular trajectory of an equatorial cross section [inset of Fig.~4(c)] and shows that the time derivative of the angular trajectories is neither flat nor jagged. Shown in Fig.~4(c) is the normalized probability distribution of persistence times from the simulated data.  The distribution fits a gamma distribution with shape and scale parameters $k = 1.9 \pm 0.6$ and $\theta = 4.0 \pm 0.9$.  These values agree with the shape and scale parameters for $20$ $\mu$m diameter droplets from the experiment.

Next, we go beyond the regime probed by our experiments, extending our analysis to a larger phase space in $\hat{U}_0$ and $\varphi$.  We find that the collective behavior of the rods depends strongly on $\hat{U}_0$; at high $\hat{U}_0$, the Yukawa amplitude is so strong that rods are locked in position, and as $\hat{U}_0 \rightarrow 0$ rods are moving independently. To find the persistence times from the simulation, $\hat{\tau}_P$, we first project the total angular momentum of the system onto the $z$-axis using the positions and velocities of each rod, $L_z(t) = \sum \mathbf{l_i}\cdot \mathbf{e}_z$, where $i$ runs from 1 to $n$, the total number of self-propelled rods on the spheroid.  The persistence time, $\hat{\tau}_P$, is the time between successive zero crossings in $L_z(t)$.  Note that this analysis differs slightly from that of the experiments. Shown in Fig.~4(c) is a phase portrait of the average persistence times from the simulation in the $\varphi$-$\hat{U}_0$ plane. The average persistence time increases as the packing fraction increases.  Physically, as the packing fraction approaches random close packing, $\hat{\tau}_P$, it becomes difficult for individual rods to displace their nearest neighbors.  This leads to a slower precession of the system's total angular momentum.


Flow reversals arise due to the precession of the overall polarization field within the shell, \textit{i.e.}, islands of aligned bacteria with large positive or negative angular momentum that move dynamically in and out of the focal plane in our experiments and the simulated cross sectional data presented in Fig.~4(b) and (c). To see this, we used a equirectangular stereographic projection which maps the azimuthal and polar coordinates $\phi$ and $\theta$ to $x$ and $y$, and given by $x = R(\phi - \pi), y = R(\pi/2 - |\theta|)$, where we have assumed that the spheroid axes $a \approx b \approx R$. Fig.~5(a) shows a schematic of the map, and in  Fig.~5(b) and (c), we see islands of aligned rods flowing counterclockwise and clockwise respectively.  Note, as $y \rightarrow \pm 15 \mu$m, rods appear to move faster due to the projection. This increased speed is equivalent to the stretching by the equirectangular projection.  Critically, the substrate geometry permits motion in the out-of-plane direction creating, in effect, sources and sinks for polarization.  In 2D, these reversals would be less probable because polarization reversals would be principally generated by contact between neighboring rods. 

Fig.~5(d) shows a simplified physical schematic of the stereographic projection of length $2\pi R$ and thickness $\delta y$. Right pointing arrows represent the populations of self propelled rods moving counterclockwise on average, and left pointing arrows represent populations of self propelled rods moving clockwise on average.  Motion can be coarse grained into average fluxes moving in the $x$ and $y$ directions with $\langle \rho^+v^+_x\rangle$ and $\langle \rho^-v^-_x\rangle$ representing counterclockwise and clockwise populations in the $\theta$ direction and $\langle \rho^+v^+_y\rangle$ and $\langle \rho^-v^-_y\rangle$ representing counterclockwise and clockwise populations in the $y$ direction.  If  $\langle \rho^+v^+_x\rangle$ dominates, the flow is polarized strongly in the counterclockwise direction; however, if  $\langle \rho^+v^+_y\rangle < 0$ and/or $\langle \rho^-v^-_y\rangle > 0$ the state can rapidly switch to a clockwise polarization. 

{In our experiment, by observing planes at constant height, we sample the large-scale precessing zonal flow structures temporally.  We observe that the average persistence times of the oscillatory motion grows with the radius of the droplet [see Fig.~3(b)].  A plausible physical explanation for this follows by recognizing that to switch flow directions, a critical number of bacteria, $N_c$, flowing clockwise must replace bacteria flowing counterclockwise. As the radius of curvature increases, it is relatively slower and more difficult to generate a switch because $N_c$ increases.  This requires more bacteria to collectively coordinate their motion and move into the optical plane. In contrast, in the limit of small radius of curvature, if a strong persistent flow is established, it is relatively easily reversed because $N_c$ is small.  Small number fluctuations of bacteria moving in the opposite direction are sufficient to switch the flow direction.}


Our study has probed the collective effects of polar active particles confined to a 2D curved, periodic spheroidal domain.  Due to out-of-plane motion, we find that azimuthal flows of the bacterial suspension oscillate dynamically with persistence time statistics that fit a $\Gamma$ distribution and depend on the double emulsion size. These results were corroborated and extended using numerical agent-based simulation.  The physical origin of these flow reversals arises from exchanging bacteria of a given polarity with bacteria of the opposite polarity. Together, our work constitutes an early step towards exploring how active systems couple to curved substrate geometries.

\acknowledgments{We thank A.~Nagarkar, P.~Ellis, S.~G.~Prasath, Y.~Jung, J.~Werner (Boston U), and  D.~Lee (Penn) for discussions.  J.~S.~Y.~thanks K.~Fahrner, who taught him how to work with \textit{E. coli}. S.G. acknowledges the Gordon and Betty Moore Foundation for support as a Physics of Living Systems Fellow through Grant No. GBMF4513.  We acknowledge partial support from grants NSF DMR 14-20570 MRSEC, NSF Simons Grant 17-64269, the Simons Foundation and the Henri Seydoux Fund.}


\end{document}